%% file: paper.tex
\documentstyle[epsf,agupp]{article}
\lefthead{Gr\"UN ET AL.}

\righthead{
Techniques for Galactic Dust Measurements in the Heliosphere
}
\received{}
\revised{}
\accepted{}

\paperid{Ms\# 99.0050}
\cpright{AGU}{}
\ccc{}

\authoraddr{E. Gr\"un, 
Max-Planck-Institut f\"ur Kernphysik, 
Postfach 10 39 80,
69029 Heidelberg,
Germany (e-mail: Eberhard.Gruen@mpi-hd.mpg.de)}

\slugcomment{To appear in the {\it Journal of Geophysical 
Research}, 1999.}
\input{update.tex}

\begin{document}

\title{Techniques for Galactic Dust Measurements in the Heliosphere
}

\author{Eberhard Gr\"un\altaffilmark{1}, 
        Markus Landgraf\altaffilmark{2},
        Mihaly Hor\'any\altaffilmark{3},    
        Jochen Kissel\altaffilmark{4},
        Harald Kr\"uger\altaffilmark{1},
        Ralf Srama\altaffilmark{1}, 
        H{\aa}kan Svedhem\altaffilmark{5},
        Peter Withnell\altaffilmark{3}}

\altaffiltext{1}{Max-Planck-Institut f\"ur Kernphysik, 
                 Postfach 10 39 80, 69029 Heidelberg, Germany} 
\altaffiltext{2}{NASA Johnson Space Center, Houston, TX\,77058, USA}
\altaffiltext{3}{Laboratory for Atmospheric and Space Physics, 
                 University of Colorado, Boulder, CO\,80309, USA}
\altaffiltext{4}{Max-Planck-Institut f\"ur Extraterrestrische Physik,
                 85740 Garching, Germany}
\altaffiltext{5}{European Space Research and Technology Centre, 
                 2200 AG Noordwijk, The Netherlands}

\begin{abstract}

Galactic interstellar dust (ISD) is the major ingredient in planetary formation. However, 
information on this important material has been extremely limited. Recently the Ulysses dust 
detector has identified and measured interstellar dust outside 1.8~AU from the Sun at ecliptic 
latitudes above $50^{\circ}$. Inside this distance it could not reliably distinguish interstellar 
from interplanetary dust. Modeling the Ulysses data suggests that up to 30\,\% of dust flux with 
masses above $10^{-16}\rm \, kg$ at 1~AU is of interstellar origin. From the Hiten satellite in high 
eccentric orbit about the Earth there are indications that ISD indeed reaches the Earth's orbit. 
Two new missions carrying dust detectors, Cassini and Stardust, will greatly increase our 
observational knowledge. In this paper we briefly review instruments used on these missions 
and compare their capabilities. The Stardust mission [{\em Brownlee et al.}, 1996] will analyze the 
local interstellar dust population by an in-situ chemical analyzer and collect ISD between 2 
and 3~AU from the Sun. The dust analyzer on the Cassini mission will determine the 
interstellar dust flux outside Venus' orbit and will provide also some compositional 
information. Techniques to identify the ISD flux levels at 1~AU are described that can 
quantify the interstellar dust flux in high-Earth orbit (outside the debris belts) and provide 
chemical composition information of galactic dust. 

\end{abstract}

\begin{article}
%=========================================================================

\section{Introduction}

Most material contained in the Earth and the other planets has resided in galactic interstellar 
dust grains (ISD) $5\cdot 10^{9}$ years ago before it was mixed and altered during the planetary 
formation process. However, information on this galactic dust has been extremely limited. 
ISD is astronomically recognized by the missing light from distant stars. Extinction of 
starlight, especially in the UV, provides a mean to quantify the amount of dust between the 
observer and the star (e.g. Mathis, 1990). Typically this dimming of starlight is observed 
averaged over galactic (kiloparsecs) distances. More locally, the existence of dust is indirectly 
inferred from the depletion of elements heavier than helium in the interstellar gas if it is 
compared to an assumed "cosmic" abundance (e.g. {\em Frisch}, 1981). 

Galactic dust is believed to originate in a variety of different stars and stellar phenomena: e.g. 
carbon-rich stars, red giants, or supernovae, all of which provide dust with characteristic but 
different chemical and isotopic signatures that gets modified during its passage through 
interstellar space [{\em Dorschner and Henning}, 1995]. A variety of presolar grains have been 
identified in primitive meteorites: e.g. diamonds, graphite, silicon carbide, or corundum grains 
[{\em Zinner}, 1998]. The identified grains are only a minute fraction of the total material 
that went 
into the protoplanetary disk. The composition of the bulk of the grains is largely unknown. 

Galactic dust grains passing through the planetary system have been detected by the dust 
detector onboard the Ulysses spacecraft [{\em Gr\"un et al.}, 1993]. Once it became evident that 
galactic dust is accessible to in-situ detection and even sample return to Earth, NASA selected 
the Stardust mission to analyze and return to Earth samples of ISD collected at asteroid belt 
distances [{\em Brownlee et al.}, 1996]. 

The following key questions need to be addressed by future in situ ISD investigations:
1. The ultimate goal of the study of interstellar grains is the determination of their elemental 
and isotopic composition (information that cannot be obtained from astronomical 
observations) in order to derive information on their sources and processes by which they 
were formed and on the conditions during their passage through interstellar space. In addition 
to accomplishing this astrophysical goal the determination of the chemical composition will 
help us to identify interstellar dust and to distinguish it from interplanetary dust. 
2. The size distribution of interstellar dust observed by spacecraft over the widest range 
possible will allow us to narrow down the size gap between spacecraft and radar 
measurements of ISD (see below). Thereby, the transition may be observed of dust that is bound to the 
local gas cloud to dust that is bound to much larger structures in the diffuse interstellar 
medium or that is even arriving directly from its source region.

We here briefly review what is known about ISD in the planetary system, and then describe 
how one can further study this fundamental material.

\section{Interstellar dust in the solar system}

Ulysses observations provided unique identification of ISD by three characteristics: 1. At 
Jupiter's distance the grains appeared to move on retrograde trajectories opposite to orbits of 
most interplanetary grains and the flow direction coincided with that of interstellar gas 
[{\em Witte et al.}, 1993], 2. A constant flux has been observed at all latitudes above the 
ecliptic plane, 
while interplanetary dust displays a strong concentration towards the ecliptic, and 3. The 
measured speeds of the interstellar grains (despite of their substantial uncertainties) were high 
(in excess of the escape speed from the solar system) which indicated orbits unbound to the 
solar system, even if one neglects radiation pressure effects. 

It has been found that interstellar gas flows through the planetary system with a speed of 
$ 26\, \rm km\, s^{-1}$ from the direction of $253^{\circ}$ ecliptic longitude 
and $5^{\circ}$ ecliptic latitude 
[{\em Lallement}, 1993, {\em Witte et al.}, 1993]. The flow of ISD was within about 
$10^{\circ}$ of that of 
interstellar gas and persisted during Ulysses' tour over the poles of the Sun where very few 
interplanetary dust is expected. ISD was identified as close as 1.8~AU from the Sun at ecliptic 
latitudes above $50^{\circ}$ [{\em Gr\"un et al.}, 1997]. 

Measurements in the ecliptic plane by Galileo confirmed that outside about 3~AU interstellar 
dust flux exceeds the flux of interplanetary grains [{\em Baguhl et al.}, 1995]. Inside the distance of 
1.8~AU Ulysses could not easily distinguish ISD from interplanetary micrometeoroids. From 
the Hiten satellite in high eccentric orbit about the Earth, however, there are indications that 
ISD indeed reaches the Earth's orbit [{\em Svedhem et al.}, 1996]. The Hiten dust detector measured 
an excess flux from the upstream interstellar flow direction. Modeling the Galileo and Ulysses 
data suggests that up to 30\,\% of dust flux with masses above $10^{-16}\rm \, kg$ at 1~AU is of 
interstellar origin [{\em Gr\"un et al.}, 1997].

Interstellar grains observed by Ulysses and Galileo range from $ 10^{-18}\rm \, kg$ to above 
$10^{-13}\rm \,  kg$ (Figure 1, see also {\em Landgraf et al.}, 1999). 
The deficiency of measured small grain 
masses is not solely caused by the detection threshold of the instrument but it indicates a 
depletion of small interstellar grains in the heliosphere. Estimates of the filtering of 0.1 
micron-sized and smaller electrically charged grains in the heliospheric bow shock region and 
in the heliosphere itself [{\em Frisch et al.}, 1999] show that these particles are strongly impeded 
from entering the planetary system by the interaction with the ambient magnetic field. Figure 
2 shows the modeled flux [{\em Landgraf}, 1999] of ISD at Earth. The model takes into account 
solar gravity and radiation pressure (characterized by the ratio, $\beta$, of radiation 
pressure over 
solar gravity), as well as, electromagnetic interaction of electrically charged dust grains 
(described by the charge-to-mass ratio, Q/m) with the interplanetary magnetic field that varies 
with the solar cycle. The annual modulation is caused by Earth's orbit around the Sun. 

More recently, even bigger ($10^{-10}\rm \, kg$) interstellar meteors have been reliably 
identified by their hyperbolic speed at 1~AU [{\em Taylor et al.}, 1996, {\em Baggaley}, 1999]. 
It is found that the mass 
distribution only overlaps with the bigger masses of the "classical" {\em Mathis, Rumpel, and 
Nordsiek} [1977] distribution of astronomically observed interstellar grains but extends to 
much larger particles. The flow direction of big particles varies over a much wider angular 
range than that of small grains observed by Ulysses and Galileo. The total mass flux of the 
Ulysses particles corresponds to the dust mass density (at $26\, \rm km\, s^{-1}$ relative 
speed) one would 
expect from observations of the gas density in the local interstellar medium 
($ \sim 5\cdot 10^{-25}\, \rm g\,cm^{-3}$) and assuming a standard cosmic abundance of elements 
(i.e. the dust mass density 
is about 1\,\% of the gas mass density, for a detailed discussion see {\em Frisch et al.}, 1999). 

Most micron-sized dust particles found by Ulysses and Galileo are bigger than the typical 
interstellar grains. The existence of a significant number of big particles in the diffuse 
interstellar medium has profound consequences for the evolution of interstellar material. The 
mass in big grains provides a significant collisional reservoir for smaller particles that would 
otherwise rapidly be depleted [{\em Gr\"un and Landgraf}, 1999].

\section{Measurements, expected and potential}

To accomplish the scientific goals, namely, to analyze the size and compositional distribution 
of interstellar grains, one must first unambiguously distinguish ISD from interplanetary dust 
and from man made debris (in near-Earth environment). Since ISD contributes only about 
30\,\% of the total sub-micron sized dust flux at 1~AU distance from the Sun, additional 
parameters have to be determined with sufficient accuracy. The flux of small interplanetary 
dust grains onto a spacecraft on a circular heliocentric orbit is constant - independent of 
spacecraft position around the Sun - whereas ISD flows from a specific direction through the 
solar system. Therefore, the direction from which ISD is detected will display variations along 
the spacecraft orbit around the Sun. To distinguish interstellar from interplanetary dust, 
measurements of the flux direction are required. Also the relative speed of interstellar grains 
will be modulated by the spacecraft's motion around the Sun, therefore, the impact speed will 
display large (about factor of 10) variations during a complete orbit. If the impact speed is 
measured with sufficient accuracy then this provides an additional criterion for interstellar 
dust identification. In addition, the determination of the dust velocity provides information on 
the heliospheric ISD dynamics, like radiation pressure and electromagnetic effects which 
modify ISD trajectories through the solar system.

The size distribution of ISD needs to be determined over the widest possible size range. The 
smallest interstellar grains that reach the inner planetary system can be as small as 0.1 micron 
in radius. Only impact ionization dust detectors are able to reliably detect and characterize 
these small grains. Also the biggest interstellar particles accessible (several microns in radius 
and above) are of interest because they are the target of sample return missions.

New compositional analyses of ISD passing through the planetary system are expected from 
the Cassini and Stardust missions. Cassini with its Cosmic Dust Analyzer (CDA) was 
launched in October 1997 and commenced dust measurements in March 1999 on approach to 
its final flyby of Venus. It will continue interplanetary and interstellar dust measurements 
until its arrival at Saturn in 2004. The Cassini CDA combines a high sensitivity dust detector 
with a mass analyzer for impact-generated ions. CDA is an impact ionization dust detector of 
$0.1\, \rm  m^2$ sensitive area for the determination of physical properties: flux, mass and speed 
distribution, electrical charge, and coarse chemical analysis ($\rm M/\Delta M = 20$ to 50). 

The Stardust Discovery mission will collect dust from the coma of Comet P/Wild 2 and 
interstellar grains and return them to Earth [{\em Brownlee et al.}, 1996]. Several times during its 
eccentric orbit about the Sun (out to about 3~AU) interstellar dust and dust from Comet Wild 
2 will be captured by impact into aerogel and brought back to the Earth in 2006. In addition, 
in situ detection and compositional measurements of cometary and interstellar grains are 
performed since April 1999 by the Cometary and Interstellar Dust Analyzer (CIDA) and dust 
flux monitors that are attached to the front shield of the spacecraft. CIDA  is a high resolution 
($\rm M/\Delta M > 100$) impact ionization mass spectrometer for chemical analysis of ISD. It has 
$ 0.009 \, \rm m^2$ sensitive area. 

Dust detectors with large sensitive area (ca. $\rm 1\, m^{2}$) but still high sensitivity 
($10^{-16}\rm \, kg$) 
need to be used to measure ISD over a wide size distribution. A Deployable Dust Detector 
System, $\rm D^{3}S$ is under development [{\em Hor\'anyi et al.}, 1998]. 
It consists of an impact detector 
of the type flown on VEGA [{\em Simpson and Tuzzolino}, 1985] and trajectory analyzer of 
charged micron sized dust grains similar to the charge measurement of Cassini CDA. In the 
following paragraphs we describe in some detail the Cassini CDA, Stardust CIDA, and the 
$\rm D^{3}S$ detector. 

\subsection{Cosmic Dust Analyzer, CDA}

The Cosmic-Dust-Analyzer measures the mass, coarse composition, electric charge, speed, 
and flight direction of individual dust particles [{\em Srama et al.}, 1996]. The Cosmic-Dust-
Analyzer, CDA, has significant inheritance from former developed space instrumentation for 
the VEGA, Giotto, Galileo, and Ulysses missions. It measures impacts from as low as one 
impact per month up to $10^{4}$ impacts per second. The instrument weighs 17~kg and 
consumes 12 Watts, the integrated time-of-flight mass spectrometer has a mass resolution 
above 20. The data transmission rate is 524 bits per second and an own pointing platform allows 
articulation of the complete instrument by up to $270^{\circ}$ around one axis.

The detection of dust particle impacts is accomplished by two different methods: (1) The Dust 
Analyzer (DA) uses impact ionization for particle detection. DA measures the electric charge 
carried by dust particles, the impact direction, the impact speed, mass and chemical 
composition. (2) The High Rate Detection system (HRD) uses two separate PVDF sensors 
(polyvinylidene fluoride, {\em Simpson and Tuzzolino}, 1985, {\em Tuzzolino}, 1992), for the 
determination of high impact rates during Saturnian ring plane crossings.

Figure 3 shows a schematic cross section of the DA sensor with its charge sensing grids, and 
electrodes. The grid system in the front provides the measurement of the dust charge and of 
components of the velocity vector. A charged dust particle entering the sensor induces a 
charge which corresponds directly to its own charge. The output voltage of the amplifier rises 
until the particle passes through the second grid. As long as the particle is located between the 
second and third grid the output voltage remains more or less constant. As soon as the dust 
particle has passed the third grid, the voltage begins to fall until the fourth grid is passed. Due 
to the inclination of $9^{\circ}$ for the inner two grids, the path length between the grids 
depends on the angle of incidence, and allows a determination of the flux direction of charged 
dust particles in one plane. The choice of $9^{\circ}$ is a compromise between angular 
resolution and the length of the instrument. The detection of particle charges as low as 
$10^{-15}\, \rm  C$ will be possible.

Hypervelocity dust impacts onto the target produce an impact plasma, i.e. a cloud comprised 
of neutral atoms, ions and electrons. Electrical charges produced by impacts onto the big gold 
plated impact ionization target are collected on the target and on the negatively biased ion 
collector. Mass and speed can be derived from the measured signals through empirical 
calibration (cf. {\em Gr\"un et al.}, 1995). Positive ions produced by impacts onto the chemical 
analyzer target (CAT) will be mass analyzed. The strong electric field between the small 
rhodium target and the grid separates very quickly impact charges of different polarity and 
accelerates the positive ions towards the multiplier. The curved shapes of the target and grid 
focus the ions onto the multiplier. This time-of-flight mass spectrometer has a flight path 
length of 0.23 m and provides elemental composition of the micrometeoroids with a mass 
resolution $\rm M/\Delta M \geq 20$ [{\em Ratcliff et al.}, 1992]. 
The signals at the output of the electron 
multiplier are sampled and digitized at 100 MHz rate and have a dynamic range of $10^{6}$. 

The measurable particle mass ranges are about $10^{-19}$ to $10^{-13}\rm \,  kg$ for 
$40\rm\, km\,s^{-1}$ 
impact speed and $10^{-16}\rm \, kg$ to $10^{-10}\,\rm kg$  for $5\, \rm km\,s^{-1}$ 
impact speed. The detection 
threshold for the impact speed is about $ 1\, \rm km\,s^{-1}$. There is no upper 
speed limit for the detection 
of particles, but the speed determination will be difficult beyond $ \rm 80\, km\,s^{-1}$. 
Electrical charges 
carried by the dust particles are measured from $10^{-15}$ to $5\cdot 10^{-13}\rm\, C$ 
for both, negative 
and positive charges. The trajectory of charged particles (above $10^{-14}\, \rm C$) can be measured 
with an accuracy of $2^{\circ}$ in one plane. The mass resolution of the time-of-flight (TOF) 
mass spectrometer lies between $\rm M/\Delta M = 20$ (for masses $<$ 30 atomic mass units, amu) 
and $\rm M/\Delta M = 50 $ (for masses $>$ 50 amu). 

\subsection{Cometary and Interstellar Dust Analyzer, CIDA}

The chemical composition of dust particles is measured that impact onto the sensor with 
speeds above a few km/s. In situ compositional analysis has been implemented in the Halley 
missions with great success [{\em Kissel et al.}, 1986, {\em Sagdeev et al.}, 1986]. The 
CIDA instrument 
onboard Stardust is a direct derivative of the PIA instrument flown on the GIOTTO 
spacecraft. 

Figure 4 shows a schematic view of the CIDA instrument. When a dust particle hits the solid 
silver target with a speed well above 1 km/s, solid ejecta, neutral and ionized molecules, 
electrons from the target and the projectile are emitted. Positive ions are analyzed by a TOF 
mass spectrometer. A charge sensitive amplifier measures the impact signal at the target. 
Accelerated by the electric field in front of the target, the ions travel into the drift tube of the 
TOF mass spectrometer. At the end of the drift tube an electrostatic reflector is located which 
deflects the ions onto an electron multiplier, and at the same time compensates the spread in 
flight times due to different initial starting energies the ions might have. Amplifiers connected 
to the multiplier allow the measurement of the time of flight spectrum. The bias at the target 
and the multiplier front stage are -1 kV and +1.3 kV, respectively. The dimensions of the 
instrument are determined by the size of the target and the desired time resolution. While the 
target size is limited by the size of the ion detector, the time resolution is mostly limited by 
the instrument's electronics. CIDA has a single stage ion reflector, followed by an open 
electron multiplier of 30 mm diameter sensitive surface area. The mass resolution of 
$\rm M/\Delta M = 250$, or time resolution $\rm t/\Delta t = 500$ is achieved by a digitization 
frequency of 80 MHz. The 
maximum detectable mass is 330 amu. The useful target size is 120 mm in diameter. For an 
impact angle of $40^{\circ}$ from the target normal, this corresponds to a sensitive area of 
$ \rm 0.009\, m^{2}$.

\subsection{Deployable Dust Detector System, $\bf D^{3}S$}

$\rm D^{3}S$ combines two existing techniques that are most readily adaptable to large surface 
area applications:  the electrostatic charge-sensing grid structure and the PVDF film. The 
proposed large surface area dust detector will apply two sets of X/Y grids for position, charge 
and velocity sensing and a PVDF film for mass determination at the bottom of the instrument 
(Figure 5). The detection principle of the charge sensing trajectory detectors is based on 
charge induction. When a charged dust grain passes close to a conducting wire, it induces a 
charge on the conductor that can be measured as a function of time. The amplitude of the 
signal is proportional to the charge on the dust particle and the duration of the signal contains 
information on the velocity of the grain in the detector. These measurements do not alter the 
charge state or the velocity vector of the dust particle [{\em Auer}, 1975; 1996; {\em 
Auer and von Bun}, 1994]. 

The detection principle of the PVDF film detectors is based on the depolarization signal a 
dust particle generates when penetrating a permanently polarized PVDF thin film [{\em Simpson 
and Tuzzolino}, 1985; {\em Tuzzolino} 1992]. Dust grains penetrating the thin PVDF film remove 
dipoles along their trajectory producing a fast electric charge pulse without requiring bias 
voltages.  The produced signal is a function of the particles mass and velocity.  PVDF film 
dust detectors have been extensively tested and calibrated in laboratory experiments and have 
an excellent track record in space experiments. PVDF sensors were flown on the VEGA 
missions to comet Halley [{\em Simpson et al.}, 1987], and are presently flown on the Stardust and 
the Cassini missions.

The charge-sensing velocity detector and the PVDF film are both well suited to large surface 
area applications. The demonstrated accuracy is $\sim 1\, \%$ in speed, $\sim 1^{\circ}$ in 
angle and $\sim 10\,\%$ in charge (for grain charges $\rm Q > 10^{-15}\, \rm C$; the accuracy improves 
for larger Q) of the charge sensing trajectory detector [{\em Auer}, 1996]. The accuracy of the 
PVDF sensor is a factor of two in mass for particles with masses $\rm m > 10^{-16}\, \rm kg$ and 
velocities $v > \rm 1\,km\, s^{-1}$ (the mass determination improves for larger $m$ and $v$; 
{\em Simpson and Tuzzolino}, 1985). Advanced features, like impact charge and mass analysis may 
be included in the future design.

\section{Outlook}

The Cassini and Stardust missions will provide new and unprecedented information on 
galactic interstellar dust passing through the solar system from the local interstellar medium. 
However, complementary information can be provided by an optimized Earth orbiting 
mission. Such a mission would combine several detectors for interstellar dust and would stay 
operational until a sufficient number of interstellar grains is analyzed. The orbit of such a 
satellite should be outside the region with high space debris contamination, i.e. above geo-
stationary altitude ($\rm > 36,000\,km$, Figure 6). The relative impact speed and corresponding flux 
of ISD onto the Earth satellite is modulated by the Earth's motion around the Sun and the 
satellite's motion around Earth. As can be seen in Fig. 6, in winter the relative speed is about 
$ 60\,\rm km\, s^{-1}$ and the ISD flux is highest. At this time optimum in-situ 
detection and analysis of 
ISD is possible. In summer the relative speed is very low and intact sample collection may be 
feasible. 

One possible future use of the above dust detector complement may be such a mission. The 
complement of the three instruments is necessary in order to meet the science requirements, 
where each individual instrument addresses a specific task. The most versatile instrument is 
CDA. It can characterize the galactic dust flux of the smallest grains that reach the Earth's 
orbit. It provides masses and speeds of individual particles, as well as some chemical 
information. CIDA is specialized on high resolution compositional measurements, whereas 
$\rm D^{3}S$ focuses on the very low flux of big galactic particles and their trajectory analysis. 

Thereby, the full range of accessible galactic grains is covered, with overlap between the 
sensitivity ranges of the different instruments. Currently, no such mission is planned, but the 
instruments described here are already developed. This would allow quick realization were an 
opportunity to arise.

\bigskip

{\bf Acknowledgments.}
We thank the anonymous referees for their valuable remarks that 
improved the paper significantly. This paper was presented at the ISSI workshop on 
"Interstellar Dust and the Heliosphere", Bern, 1998. ISSI's support for this activity is 
acknowledged.

\end{article}

\begin{figure}
\epsfxsize=9.0cm
\epsfbox{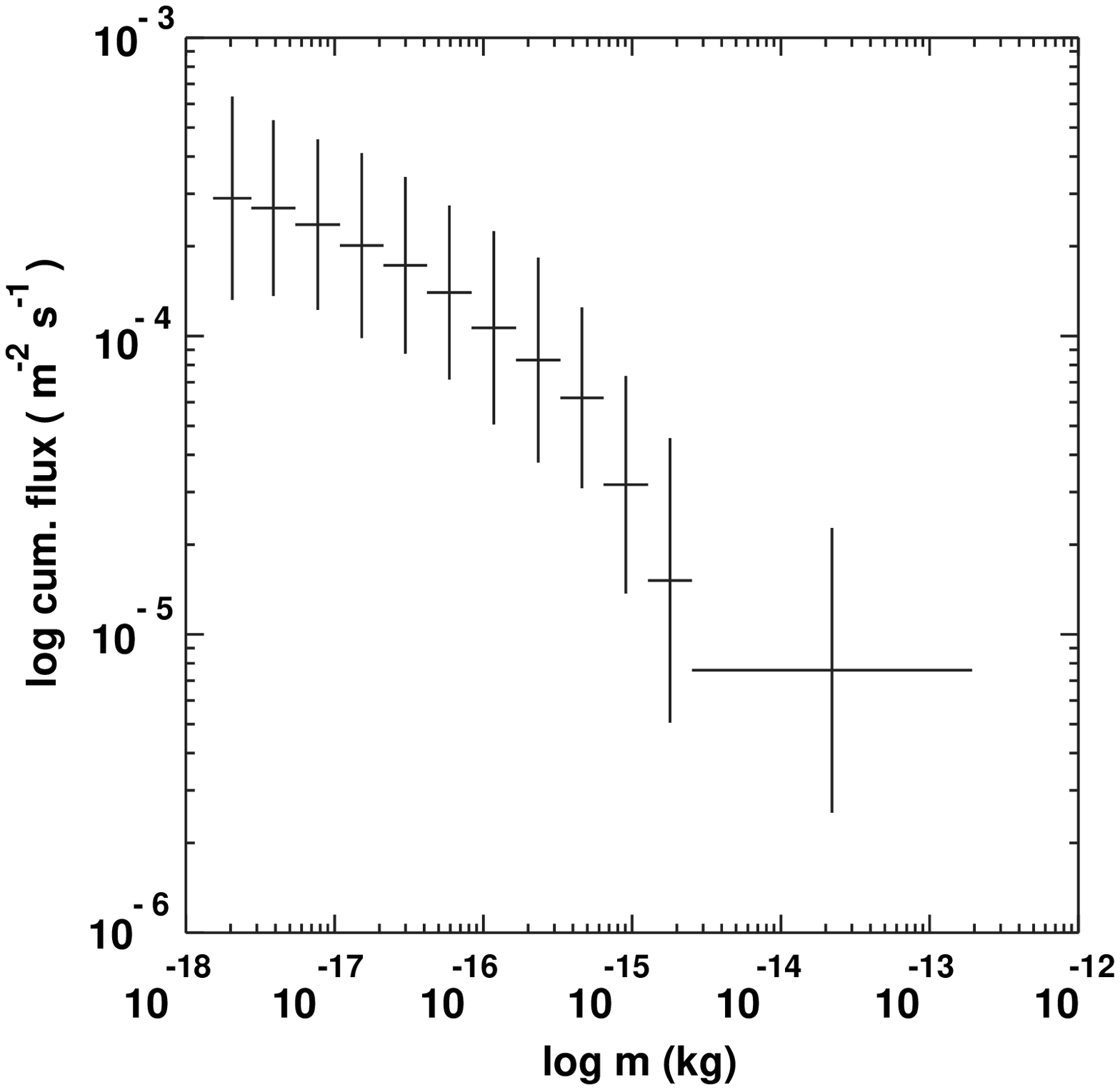}
\vspace{-3cm}
        \caption{\label{fig1}
Cumulative flux of interstellar grains observed by the Ulysses dust instruments (for 
about 3 years after Jupiter fly-by, excluding the ecliptic plane crossing, {\em Landgraf},
 1999). The 
detection threshold of the detectors is $10^{-18}\, \rm kg$ at $26\, \rm km\,s^{-1}$ impact speed.
 The radius scale 
(a) is derived from the masses by assuming spherical particles of $2500\, \rm kg\,m^{-3}$ density. 
}
\end{figure}

\begin{figure}
\epsfxsize=14.0cm
\epsfbox{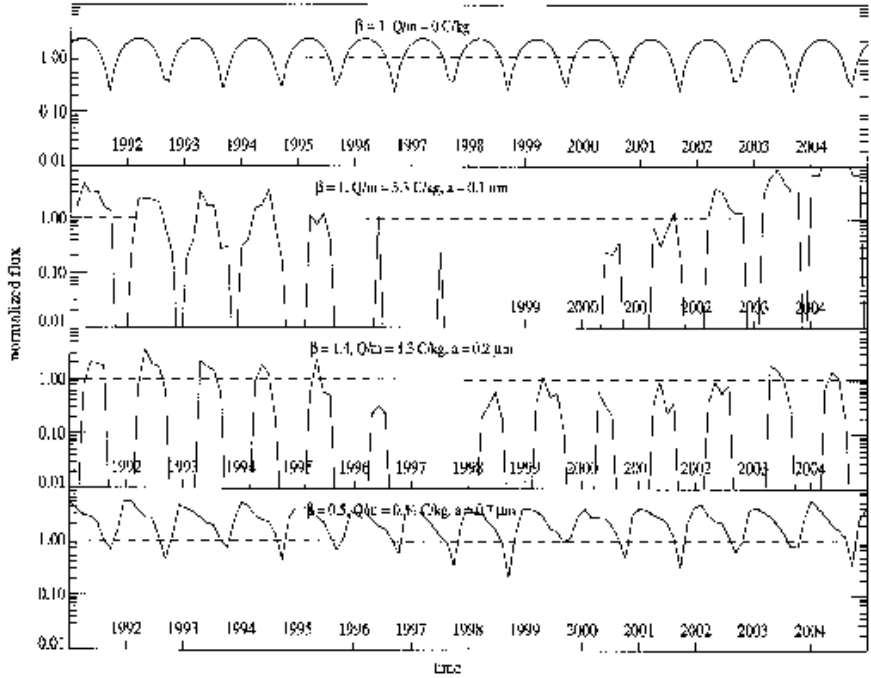}
%\vspace{-2cm}
        \caption{\label{fig2}
Model fluxes of interstellar grains of various sizes and corresponding parameters as 
function of time. Parameters are $\beta$ (ratio of radiation pressure over gravity), Q/m (charge-to-
mass ratio), and a (radius of spherical particle with density $2500\rm\, kg\,m^{-3}$). Top panel: 
radiation pressure just cancels gravity ($\beta = 1$), no charge effects; second panel: $a = 0.1\rm\, \mu m $
particle with corresponding beta and Q/m parameters; third panel: $a = 0.2\rm\, \mu m$ particle with 
corresponding $\beta$ and Q/m parameters; fourth panel: $a = 0.7\rm\, \mu m$ particle with corresponding $\beta$ 
and Q/m parameters.
}
\end{figure}

\begin{figure}
\epsfxsize=11.0cm
\epsfbox{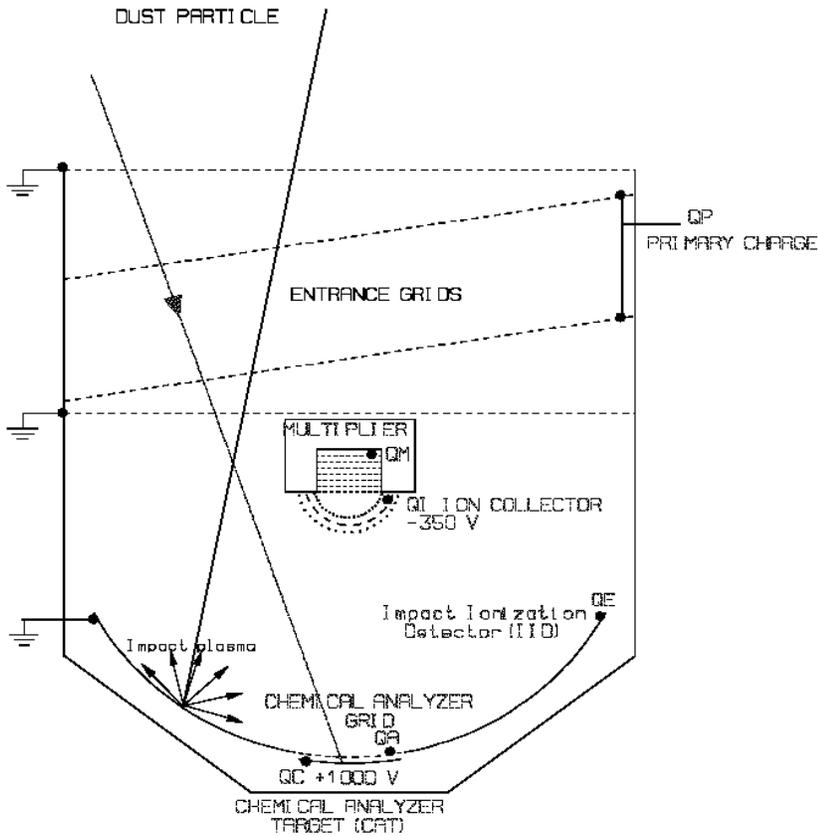}
        \caption{\label{fig3}
Schematic cross section of the Cosmic Dust Analyzer, CDA. The sensor consists of 
four charge sensing entrance grids, the hemispherical target, and the ion collector with the 
multiplier in the center. The innermost and outermost of the four entrance grids are grounded, 
the two inclined grids are connected to a charge sensitive amplifier ($Q_P$) which provides 
measurements of the induced dust particle charge and of the particle velocity. Dust particles 
(two cases are indicated) can impact either on the big gold plated impact ionization target 
(IID, diameter 0.41 m) or the small rhodium chemical analyzer target (CAT, diameter 0.16 m) 
in the center. An electric field of 350 Volts separates electrons (collected by the targets) and 
ions (collected by the ion grid). Charge sensitive amplifiers collect the charges at the two 
target electrodes ($Q_E, Q_C$) and the acceleration grid ($Q_A$). The acceleration grid is located 3 
mm in front of the target and electrically grounded whereas CAT is at a potential of +1000 
Volts. The ion collector is located in the center of the detector. Amplifiers are connected to 
the ion grid ($Q_I$), and the multiplier ($Q_M$) and provide measurements of the total ion charge 
released and the time-of-flight spectrum over a 0.23 m drift distance. The multiplier signal is 
sampled at a rate of 100 MHz. 
}
\end{figure}

\begin{figure}
\vspace{-1cm}
\epsfxsize=14.0cm
\epsfbox{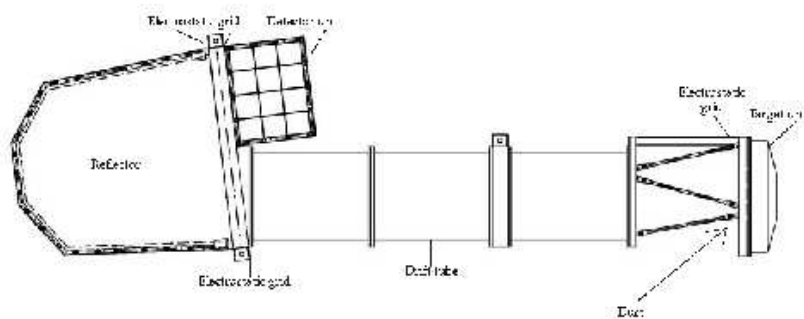}
        \caption{\label{fig4}
Cross section through the Cometary and Interstellar Dust Analyzer, CIDA. CIDA is a 
time-of-flight impact ionization mass analyzer of $\sim 0.009\rm\, m^{2}$ target size. The 
target unit houses the positively biased (+1000 V) impact target and the grounded acceleration 
grid in front of it. The open structure in front of the target assembly allows dust particles to 
reach the target. The ion drift tube of 0.55 m length is on axis with the target normal. The 
reflector unit is separated by a grid from both the drift tube and the ion detector. The 
electrostatic reflector deflects the ions onto the ion detector in such a way that ions of the 
same mass arrive at about the same time at the detector. The ion detector is a large-area open 
electron multiplier. 
}
\end{figure}

\begin{figure}
\epsfxsize=12.5cm
\vspace{-8cm}
\epsfbox{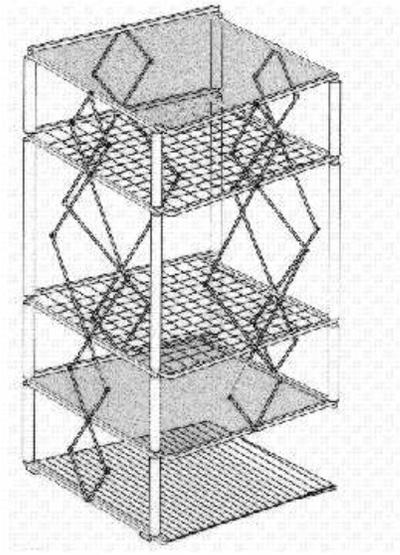}
        \caption{\label{fig5}
The Deployable Dust Detector System, $\rm D^{3}S$, is a hybrid deployable instrument, 
that consists of an electrostatic charge-sensing grid structure and the PVDF impact detector. 
The top two sensing levels are the grids of charge sensing wires, and the lower level is the 
thin PVDF film sensor. The charge sensing grids are spaced 0.83 m from each other and the 
sensor area is $1\,\rm m^{2}$ for each level. A grounded electromagnetic shield grid is mounted 
above the top charge sensing grid, and another between the lower charge sensing grid and the 
PVDF sensor. The volume between these two shields is surrounded by aluminized Kapton 
which is also grounded, forming a complete Faraday cage around the charge sensing grids. 
$\rm D^{3}S$ is composed of both rigid deployable elements and inflatable members. In the stowed 
configuration, the detector occupies a volume of 1.18 by 0.30 by 0.22 m.  The detector's mass 
is 29.4 kg.
}
\end{figure}

\begin{figure}
\epsfxsize=15.0cm
\epsfbox{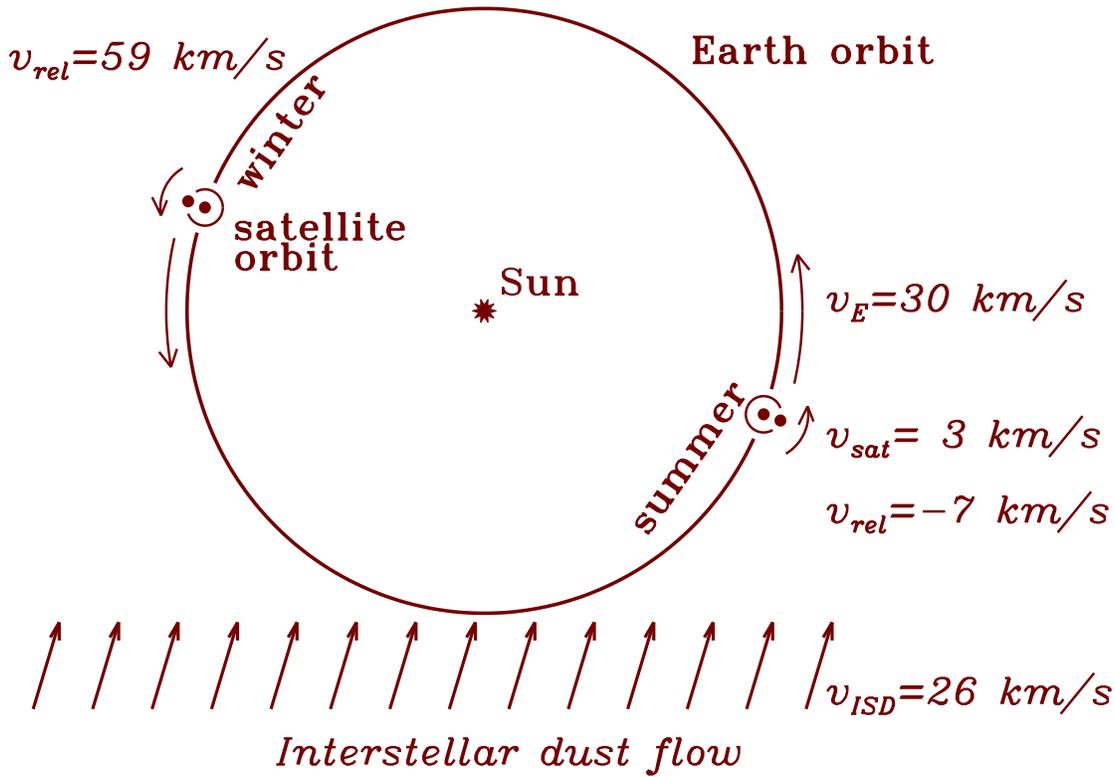}
\vspace{2cm}
        \caption{\label{fig6}
Mission scenario of an interstellar dust mission in high-Earth orbit. Shown is the orbit 
of Earth around the Sun and the direction of the interstellar gas and dust flow at speed 
$v_{\rm ISD}$. A 
satellite orbits the Earth at a speed of $v_{\rm Sat} \rm = 3\,km\,s^{-1}$ 
(orbits are not to scale). Two positions of 
the Earth (right: late summer, and left: late winter) and the satellite are shown. The 
corresponding relative speeds, $v_{\rm rel}$, to and fluxes, $F$, of interstellar grains 
are given for both 
positions. We have assumed for simplicity that radiation pressure and solar gravity for ISD 
particles just cancel each other , therefore, ISD travels on straight trajectories through the 
solar system. 
}
\end{figure}

\end{document}

%% file: update.tex
\makeatletter
\let\reset@font\empty
\makeatother